\documentclass[prd,reprint,nofootinbib,showpacs,superscriptaddress]{revtex4-1}
\usepackage{graphicx} 
\usepackage{hyperref}
\usepackage{amsfonts}
\usepackage{amsmath,amssymb}
\usepackage{bm} 
\usepackage{color}
\usepackage{epstopdf} 
\usepackage{epsfig}
\usepackage{subfig} 
\usepackage{float}

{\rm }

\def\be{\begin{equation}}
 \def\ee{\end{equation}}
 \def\bea{\begin{eqnarray}}
 \def\eea{\end{eqnarray}}
 \def\bes{\begin{eqnarray}}
 \def\ees{\end{eqnarray}}
 \def\bi{\begin{itemize}}
 \def\ei{\end{itemize}} 

\def\2{\frac{1}{2}}
\def\4{\frac{1}{4}}

\begin{document}

\title{Quantum secret sharing using weak coherent states}

\author{Warren P. Grice}
\email{Current address: Qubitekk, LLC, Vista, California 92081, USA.}
\affiliation{Quantum Information Science Group, Computational Sciences and Engineering Division, Oak Ridge National Laboratory, Oak Ridge, TN 37831, USA}

\author{Bing Qi}
\email{qib1@ornl.gov}
\affiliation{Quantum Information Science Group, Computational Sciences and Engineering Division, Oak Ridge National Laboratory, Oak Ridge, TN 37831, USA}
\affiliation{Department of Physics and Astronomy, The University of Tennessee, Knoxville, TN 37996, USA}

\date{\today}
\pacs{03.67.Dd}

\begin{abstract}

Secret sharing allows a trusted party (the dealer) to distribute a secret to a group of players, who can only access the secret cooperatively. Quantum secret sharing (QSS) protocols could provide unconditional security based on fundamental laws in physics. While the general security proof has been established recently in an entanglement-based QSS protocol, the tolerable channel loss is unfortunately rather small. Here we propose a continuous variable QSS protocol using conventional laser sources and homodyne detectors. In this protocol, a Gaussian-modulated coherent state (GMCS) prepared by one player passes through the secure stations of the other players sequentially, and each of the other players injects a locally prepared, independent GMCS into the circulating optical mode. Finally, the dealer measures both the amplitude and the phase quadratures of the receiving optical mode using double homodyne detectors. Collectively, the players can use their encoded random numbers to estimate the measurement results of the dealer and further generate a shared key. We prove the unconditional security of the proposed protocol against both eavesdroppers and dishonest players in the presence of high channel loss, and discuss various practical issues.
\footnote{This manuscript has been authored by UT-Battelle, LLC under Contract No. DE-AC05-00OR22725 with the U.S. Department of Energy. The United States Government retains and the publisher, by accepting the article for publication, acknowledges that the United States Government retains a non-exclusive, paid-up, irrevocable, world-wide license to publish or reproduce the published form of this manuscript, or allow others to do so, for United States Government purposes. The Department of Energy will provide public access to these results of federally sponsored research in accordance with the DOE Public Access Plan (http://energy.gov/downloads/doe-public-access-plan).
}

\end{abstract}

\maketitle

\section{Introduction}
\label{sec:1}

Secret sharing is a cryptographic primitive with important practical applications \cite{Shamir79, Blakley79}. In this protocol, a dealer distributes a secret message $M$ to $n$ players in such a way that at least $k\leq n$ players have to work together to decode the message. This is called a $(k,n)-$threshold scheme. In this paper, we will focus on the $(n,n)-$threshold secret sharing protocol, which means all the $n$ players have to work together to decode the dealer's message.

If the dealer shares an independent secure key $K_i$ (\textit{i}=1,2..\textit{n}) with each player and the length of each key is the same as that of the message, then a $(n,n)-$threshold scheme can be implemented as follows. The dealer first generates a new key $K=K_1\oplus K_2\oplus ...\oplus K_n$ where ``$\oplus$'' denotes addition modulo 2, then encodes message $M$ using $K$ and broadcasts the encrypted message $E=M\oplus K$. Obviously, only when the $n$ players work together can they determine $K$ and thus decode $M$ from $E$. 

The security of the above scheme relies on the security of each individual key. Two-party quantum key distribution (QKD) protocols can be employed to generate unconditional secure keys through insecure channels \cite{Gisin02, Scarani09, Lo14, Diamanti16}. The dealer could establish a QKD link with each of the players and generate $n$ individual keys before running the secure sharing protocol. However, such an implementation is rather inefficient for large $n$. Various quantum secret sharing (QSS) protocols \cite{Hillery99, Karlsson99, Tittel01, Xiao04, Chen05, Lance04,  Zhou18} have been proposed aiming at achieving both proven security and high efficiency. 

The security of QSS is deeply connected to that of QKD. Nevertheless, in contrast to a point-to-point two-party QKD protocol, a QSS protocol typically involves more participants and some of them might be dishonest. This allows additional hacking strategies and makes the security analysis of a QSS protocol more demanding than that of QKD. The application of continuous variable (CV) QKD techniques to analyze CV-QSS security was first proposed in \cite{Lau2013}. More recently, the security proof of CV-QSS against both eavesdroppers in the channels and dishonest players appeared \cite{Kogias17}. Like most other QSS protocols, the protocol studied in \cite{Kogias17} is based on multiparty quantum entanglement which may be difficult to implement with today's technology when $n$ is large. Furthermore, the tolerable channel losses presented in \cite{Kogias17} are quite small.

On another front, to ease the implementation difficulties, single qubit sequential QSS protocols have been proposed and experimentally demonstrated \cite{Schmid05}. The basic idea is shown in Fig.1. A single photon prepared in an initial polarization state propagates from party to party sequentially. Each party independently applies a random BB84-type \cite{BB84} polarization rotation on the same photon. Finally, the last recipient performs a polarization measurement. In half of the cases, the combination of the basis choices by all the parties results in a deterministic measurement result at the last recipient. These instances could be used to implement secret sharing when equipped with an appropriate postprocessing procedure. We remark that a similar design has been proposed and demonstrated in multiuser QKD \cite{Phoenix95, Grice15}. 

\begin{figure}[t]
	\includegraphics[width=.5\textwidth]{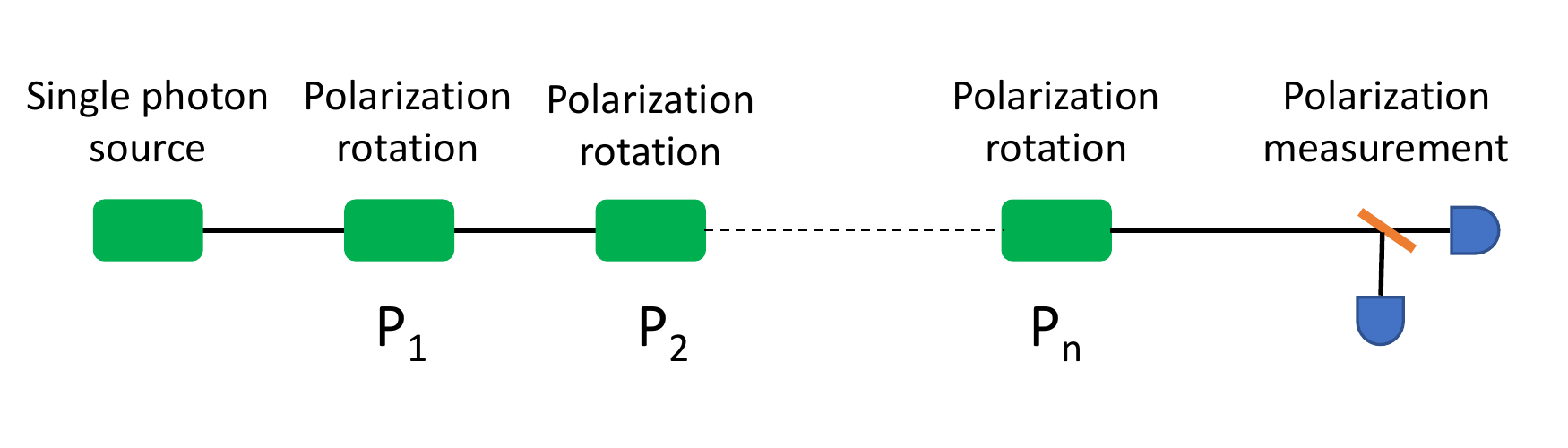}
	\captionsetup{justification=raggedright,
					singlelinecheck=false }
	\caption{Single qubit sequential QSS protocol \cite{Schmid05}.}
	\label{fig:1}
\end{figure}

While the above scheme can significantly improve the feasibility of QSS, its general security is still under debate \cite{He07, Schmid07, He10}. Furthermore such a design is vulnerable to Trojan-horse attacks where a malicious eavesdropper could send in multiphoton signals to the polarization rotation device of the targeted party and unambiguously determine the corresponding polarization rotation by measuring the output signals. We remark that in the context of QKD, a similar problem has been investigated in the so-called ``plug-and-play'' design \cite{Muller97}, where Bob sends a strong unmodulated laser pulse to Alice through an insecure channel, who in turn encodes information and sends it back to Bob after attenuating it to the single-photon level. Since the laser pulses from Bob to Alice are strong classical signals, the security issue due to the bidirectional feature of Alice's system could be mitigated by characterizing the light pulses received by Alice using conventional photodetectors \cite{zhao08, zhao10}. However, it is more difficult to apply the same countermeasures in the case of single qubit sequential QSS, where the attacker can use a weaker probe signal. This is not only because the QSS design does not employ an attenuator (as in the plug-and-play design), but also because the attacker can make use of both ports of the QSS device rather than probing and detecting via a single port.

In this paper, we will address both the security and the practicability of QSS. We propose a CV sequential QSS protocol based on conventional laser sources and homodyne detectors. The main idea is instead of modulating the quantum state of a ``passing through'' photon, each player injects a locally prepared quantum state into a circulating optical mode using a beam splitter. This prevents the eavesdroppers to access or interfere the quantum state preparation process and makes our scheme resilient to Trojan horse attacks. By choosing an appropriate beam splitting ratio, the additional loss introduced by each player's system can be extremely small, making the protocol extendable to a large number of players. Furthermore, by extending the ideas introduced in  \cite{Kogias17}, we prove the general security of the proposed protocol against both eavesdroppers and dishonest players in the presence of high channel loss.

This paper is organized as follows: In Section \ref{sec:2}, we present details of the proposed QSS scheme and provide a general security proof. In Section \ref{sec:3}, we conduct numerical simulations based on practical system parameters to show its feasibility. In Section \ref{sec:4}, we discuss various implementation issues and possible extensions.

\section{The protocol and its security}
\label{sec:2}

Inspired by the single qubit sequential QSS protocol \cite{Schmid05} and the Gaussian-modulated coherent-state (GMCS) QKD \cite{GMCSQKD}, we propose a CV-QSS protocol. As shown in Fig.2, $n$ players and the dealer are connected by a single communication channel such as a telecom fiber. For each quantum transmission, the first player $P_1$ at one end of the link prepares a coherent state $|x_1+ip_1\rangle$ and sends it to the adjacent player $P_2$. Here $x_1$ and $p_1$ are independent Gaussian random numbers with zero mean and a variance of $V_1 N_0$, where $V_1$ is the modulation variance chosen by $P_1$, and $N_0$ = 1/4 denotes the shot-noise variance. The above coherent state passes through a highly asymmetric beam splitter (with a transmittance $t_B\cong 1$) located within the secure station of $P_2$, and continues its journey to the next player. In the mean time $P_2$ locally prepares an independent GMCS and couples it into the same spatiotemporal mode as the signal from $P_1$ via the second input port of the beam splitter. By carefully controlling the modulation variances and having knowledge of the reflectivity of the asymmetric beam splitter, $P_2$ can introduce phase-space displacements of $\lbrace x_2,p_2\rbrace$. All the other players perform similar operations. At the end, the quantum state that arrives at the dealer can be described by $|\sum_{k=1}^n \sqrt{T_k}x_k+i\sum_{k=1}^n \sqrt{T_k}p_k\rangle$, where $T_k$ is the overall transmittance (including losses due to the channel and the beam splitters) experienced by the quantum signal from the $k^{th}$ player. The dealer measures both the amplitude and the phase quadratures of the received optical mode by performing double homodyne detection. Intuitively, if all the players collaborate with each other and share the encoded Gaussian random numbers, they can acquire a good estimation of the dealer's measurement results. This allows the dealer to generate a secure key which can only be known by the whole group of $n$ players and not by any subset of them. The dealer can further use the above key to implement the $(n,n)-$threshold secret sharing protocol.

The QSS protocol is summarized as below:

\begin{center}
Quantum stage
\end{center}
\begin{enumerate}
\item The first player $P_1$ draws a pair of Gaussian random numbers $\lbrace x_1, p_1\rbrace$, prepares a coherent state $|x_1+ip_1\rangle$ and sends it to the adjacent player.
\item Using a highly asymmetric beam splitter, each player down the link injects a locally prepared GMCS into the same spatiotemporal mode as the signal from $P_1$.
\item The dealer measures the amplitude and phase quadratures of the received optical mode by performing double homodyne detection. The measurement results $\lbrace x_d, p_d\rbrace$ are kept as raw data.
\item The above procedure is repeated many times to generate enough raw data. This completes the quantum stage of the protocol.

\begin{center}
Classical post-processing stage
\end{center}

\item The dealer randomly selects a subset of the raw data and requests all the players to announce the corresponding Gaussian random numbers. Combined with the corresponding measurement results, the channel transmittance $\lbrace T_1, T_2, ..., T_n\rbrace$ can be determined \cite{note1}. All the parties discard the disclosed data.
\item The dealer assumes $P_1$ is honest and all the other players are dishonest.
\item The dealer randomly selects a subset of remaining raw data and requests all the players except $P_1$ to announce their corresponding raw data.
\item The dealer displaces the measurement results of the subset in step 7 using $x_R=x_d-\sum_{k=2}^n \sqrt{T_k}x_k; p_R=p_d-\sum_{k=2}^n \sqrt{T_k}p_k$. From $\lbrace x_R, p_R\rbrace$ and $P_1$'s raw data for the same subset, the dealer and $P_1$ estimate a lower bound of secure key rate $R_1$ (in unit of bits per pulse) of two-party QKD following the standard postprocessing procedures in the GMCS QKD \cite{GMCSQKD, Diamanti15}. All the parties discard the disclosed data.
\item Steps 6-8 are repeated $n$ time. In each run, a different player is selected as the honest player. At the end, the dealer has $n$ secure key rates $\lbrace R_1, R_2,..., R_n\rbrace$. 
\item The dealer determines the secure key rate $R$ of the QSS protocol as the minimum of $\lbrace R_1, R_2,..., R_n\rbrace$, and generates the final secure key from undisclosed data using the reverse reconciliation scheme developed in GMCS QKD \cite{GMCSQKD, Diamanti15}. Note that, in reverse reconciliation, classical information goes from the dealer to the players. Accordingly, this process can be accomplished without the cooperation of the players. The dealer can implement a QSS protocol by using the final secure key to encrypt the message to be shared. Collaboratively, the $n$ players can recover the final secure key (thus the dealer's message) using their Gaussian random numbers and the classical information announced by the dealer. Any group of $n-1$ (or fewer) players can only gain an exponentially small amount of information about the final secure key.
\end{enumerate}

\begin{figure}[t]
	\includegraphics[width=.5\textwidth]{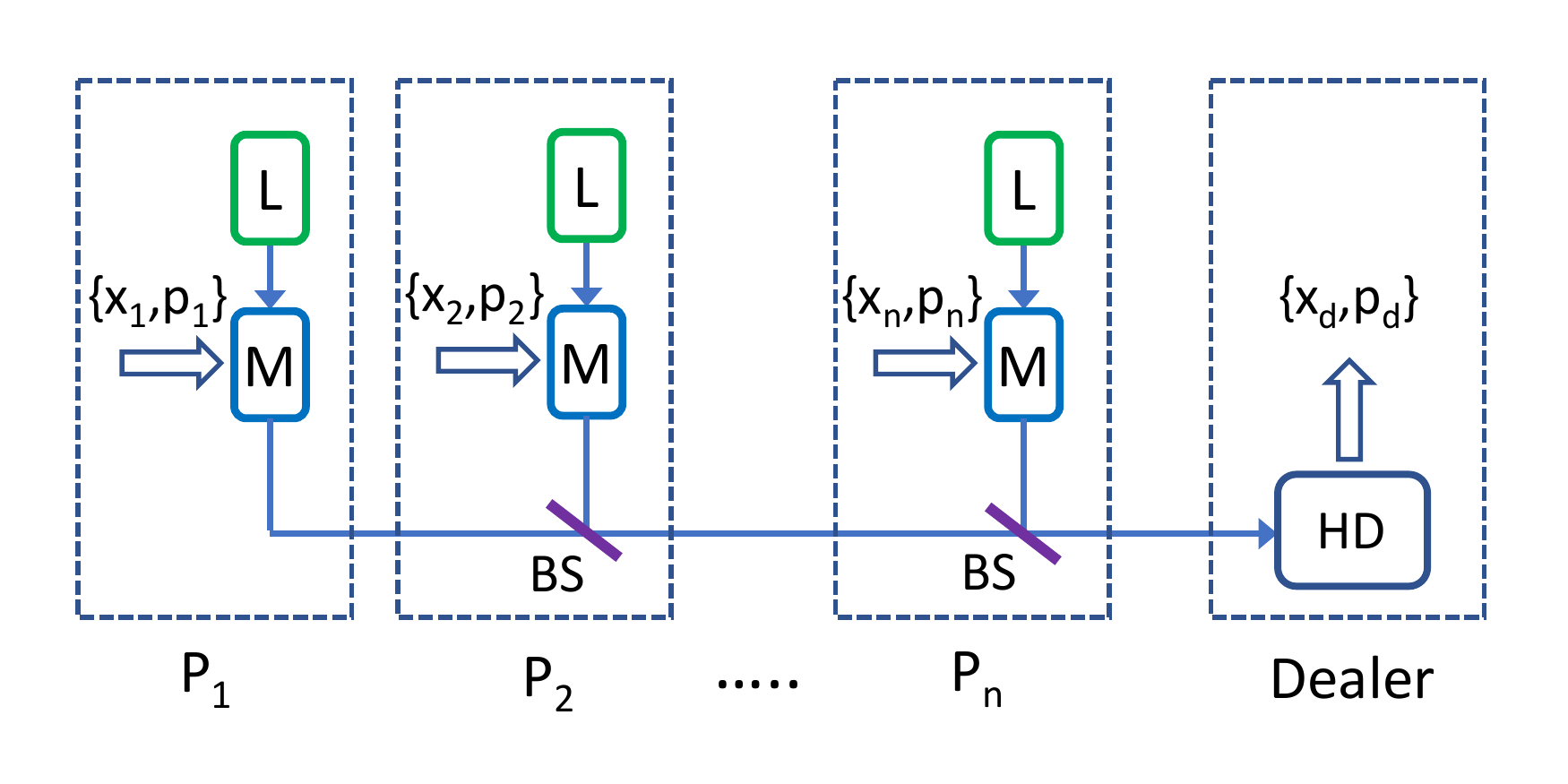}
	\captionsetup{justification=raggedright,
					singlelinecheck=false }
	\caption{The proposed CV-QSS protocol. L-laser; M-modulator; BS-beam splitter; HD-double homodyne detector}
	\label{fig:2}
\end{figure}

The data reconciliation procedure in the last step of the protocol is the same as that in the standard GMCS QKD (see \cite{Diamanti15} and references therein). Note that in the above protocol, we have implicitly assumed that all the parties share a phase reference. We will discuss how to establish such a phase reference in Section 4.

The security analysis of a QSS protocol is typically more involved than that of QKD. The general security proof against both eavesdroppers in the channels and dishonest players only appeared recently \cite{Kogias17}. In \cite{Kogias17}, the dealer prepares a multiparty continuous-variable entangled state, keeps one mode and distributes the other modes to the players. Homodyne detection is carried out by each party on the corresponding mode. One important idea in \cite{Kogias17} is to treat the measurement results announced by the players as input or output from uncharacterized devices while the dealer and the corresponding device are assumed to be trusted. This allows them to apply the tools developed in one-sided device-independent QKD \cite{Tomamichel11, Walk16} to the security analysis of QSS protocol. Nevertheless, the tolerable channel losses presented in \cite{Kogias17} are quite small.

In this paper, we follow a security proof strategy similar to that in \cite{Kogias17} by connecting the security of QSS with that of the underlying two-party QKD. In our CV-QSS protocol, the dealer needs to generate a secure key from the measurement results using reverse reconciliation. The question is at what rate the secure key can be generated (the lower bound of the secure key rate) such that only when all the $n$ players work together, can they recover the dealer's secure key, while any group of $n-1$ (or fewer) players can only gain an exponentially small amount of information. The above problem can be connected to QKD as follows: Imagine that the dealer requests a group of $n-1$ players to publicly announce their Gaussian random numbers while the last player (Alice) keeps her data private. In this case Alice (who holds the complete information of the $n$ players) should be able to recover the secure key while the $n-1$ players do not have sufficient information for key recovery. This is equivalent to a two-party QKD problem where two honest users (Alice and the dealer) try to generate a secure key against all the other $n-1$ players (and potential eavesdroppers in the channel). So the secure key rate of QSS is the same as that of QKD and can be calculated using standard security proofs of QKD (see more details in the next paragraph). Since Alice is assumed to be honest in the above picture, it is reasonable to assume the device controlled by her is also trusted. This suggests that we can use the standard security proof of QKD with trusted devices to evaluate the secure key rate. Since the secure key of QSS should be secure against \emph{any} group of $n-1$ players, the dealer needs to repeat the above procedure $n$ times: the dealer evaluates potential secure key rates of QKD with each individual player (by assuming all the other players are dishonest) and chooses the smallest one among them as the secure key rate for QSS (steps 6-9 in the protocol). This guarantees the security against the collaborating attacks between the eavesdropper and any $n-1$ (or fewer) players. By employing the security proof of standard QKD, a highly efficient, loss-tolerant QSS can be achieved. Note that we have adopted a similar security proof strategy in a recent entanglement based QSS demonstration \cite{Williams2019}.

Next we discuss how to evaluate the secure key rate of QKD between the dealer and a chosen player given that all the other $n-1$ players are dishonest. Here, we use a security argument similar to the one used in \cite{Qi15}. As specified in steps 7-8 of the protocol, after the dealer has decided which player to conduct QKD with, he (or she) requests all the other players to announce encoded random numbers for a randomly chosen subset of the raw data. The dealer then displaces the corresponding measurement results using $x_R=x_d-\sum_{k=2}^n \sqrt{T_k}x_k; p_R=p_d-\sum_{k=2}^n \sqrt{T_k}p_k$ and estimates a lower bound for the QKD key rate with the player chosen above. Since the displacement operation commutes with homodyne detection, instead of displacing the measurement results, the dealer could perform phase-space displacements before double homodyne detection. We can further assume this \emph{virtual} displacement operation is performed by the $n-1$ players outside the dealer's secure station without weakening the security of the protocol. In this picture, the actual protocol has been reduced to the standard QKD where all the operations by the other $n-1$ players (and potential eavesdroppers) are conducted in the channel before the two QKD users start the postprocessing process. Thus the standard security proof of the GMCS QKD can be applied. Note that the above security analysis covers the cases when the $n-1$ players do not execute the protocol honestly.

Note that in this paper, we have assumed that the dealer performs homodyne detection while the players prepare quantum states. In this scenario, the homodyne detector can be trusted and this allows us to apply the standard security proof of CV-QKD. Furthermore, we can apply the trusted detector noise model by assuming that both the detector efficiency and the detector noise are well calibrated and out of the adversary's control. This approach can typically lead to a better QKD performance and has been widely adopted in long-distance CV-QKD experiments \cite{GMCSQKD, Lodewyck07,Qi07, Jouguet13, Huang16, ZLC17}. We will discuss other possible arrangements in Section 4.
 
To evaluate the performance of the proposed QSS protocol, in the next section we conduct numerical simulations based on realistic system parameters.

\section{Numerical simulations}
\label{sec:3}

We assume that the quantum channel is telecom fiber with an attenuation coefficient of $\gamma$. Numerical simulations are conducted based on a specific configuration: the fiber length between the dealer (Bob) and the farthest player (Alice) is $L$. All the other $n-1$ players are distributed between them with equal separation. According to step 10 in the protocol, the secure key rate of the QSS protocol is the smallest secure key rate of two-party QKD evaluated between the dealer and each player. Given that each player introduces the same amount of excess noise (defined as $\varepsilon_0$ in the shot-noise unit), the smallest QKD key rate under normal operation will be the one between Alice and Bob. This is the key rate we will evaluate below. Note that to implement the proposed protocol in practice, the dealer should evaluate a secure key rate with each player using experimental data, and choose the smallest one as the secure key rate for QSS.

The asymptotic secure key rate of two-party GMCS QKD, in the case of reverse reconciliation, is given by Refs.~\cite{Lodewyck07, Fossier09} 
\bes\label{eq1} R=fI_{AB}-\chi_{BE}, \ees
where $I_{AB}$ is the Shannon mutual information between Alice and Bob; $f$ is the efficiency of the reconciliation algorithm; $\chi_{BE}$ is the Holevo bound between Eve (including external eavesdroppers and the other $n-1$ players) and Bob. $I_{AB}$ and $\chi_{BE}$ can be determined from the channel loss, observed noises, and other QKD system parameters. Note that all the noise terms in this paper are defined in shot-noise units.

The channel transmittance of the $k^{th}$ player is given by
\bes\label{eq2} T_k=10^{\frac{-\gamma l_k}{10}},\ees
where $l_k=\frac{n-k+1}{n}L$ is the fiber length between the dealer and the $k^{th}$ player. Here without compromising the practicability,  we have assumed that the transmittance of the beam splitter at each player's station is $t_B\cong 1$. 

The excess noise contributed by the $k^{th}$ player, when referred to the channel input, is given by
\bes\label{eq3} \varepsilon_k=\dfrac{T_k}{T_1}\varepsilon_0.\ees  

Note that the excess noise is defined as the additional noise above the vacuum noise associated with non-unity channel transmittance. Under normal operation (no eavesdroppers in the channel), the excess noise is mainly due to system imperfections, such as detector noise, errors in quantum state preparation, background light, etc.  In (3), $\varepsilon_0$ is defined as the variance of the excess noise from each player. Since the secure key rate given below is estimated using noises referred to the channel input (at Alice), we calculate the excess noise from the $k^{th}$ player by dividing $\varepsilon_0$ by the transmittance from Alice to the $k^{th}$ player $\frac{T_1}{T_k}$.   

In the case of conjugate homodyne detection, the noise added by Bob's detector (referred to Bob's input) is given by \cite{Fossier09}
\bes\label{eq4} \chi_{het}=[1+(1-\eta_D)+2\upsilon_{el}]/\eta_D,\ees
where $\eta_D$ and $\upsilon_{el}$ are the efficiency and noise variance of Bob's detector.

The channel-added noise referred to the channel input is given by
\bes\label{eq10} \chi_{line}=\frac{1}{T_1}-1+\sum_{k=1}^n \varepsilon_k,\ees
where the term $\frac{1}{T_1}-1$ represents vacuum noise due to the channel loss.

The overall noise referred to the channel input is given by
\bes\label{eq5} \chi_{tot}=\chi_{line}+\dfrac{\chi_{het}}{T_1}.\ees

Since both quadratures can be used to generate the secure key, the mutual information between Alice and Bob is given by
\bes\label{eq6} I_{AB}=\log_2\dfrac{V+\chi_{tot}}{1+\chi_{tot}}, \ees
where $V=V_A+1$, and $V_A$ is Alice's modulation variance.

To estimate $\chi_{BE}$, we adopt the realistic noise model where loss and noise of Bob's detector are assumed to be trusted and cannot be accessed by the eavesdropper \cite{GMCSQKD, Lodewyck07,Qi07, Jouguet13, Huang16, ZLC17}. Under this model, $\chi_{BE}$ is given by Ref.~\cite{Lodewyck07} 
\bes\label{eq7} \chi_{BE}=\sum_{i=1}^2 G\left( \dfrac{\lambda_i-1}{2} \right)  - \sum_{i=3}^5 G\left( \dfrac{\lambda_i-1}{2}\right),  \ees
where $G(x)=(x+1){\rm{log}}_2(x+1)-x{\rm{log}}_2x$.

\bes\label{eq8} \lambda_{1,2}^2=\frac{1}{2} \left[ A\pm \sqrt{A^2-4B} \right], \ees
where
\bes\label{eq9} A=V^2 (1-2T_1)+2T_1+T_1^2 (V+\chi_{line})^2, \ees
\bes\label{eq10}B=T_1^2(V\chi_{line}+1)^2. \ees

\begin{figure}[t]
	\includegraphics[width=.5\textwidth]{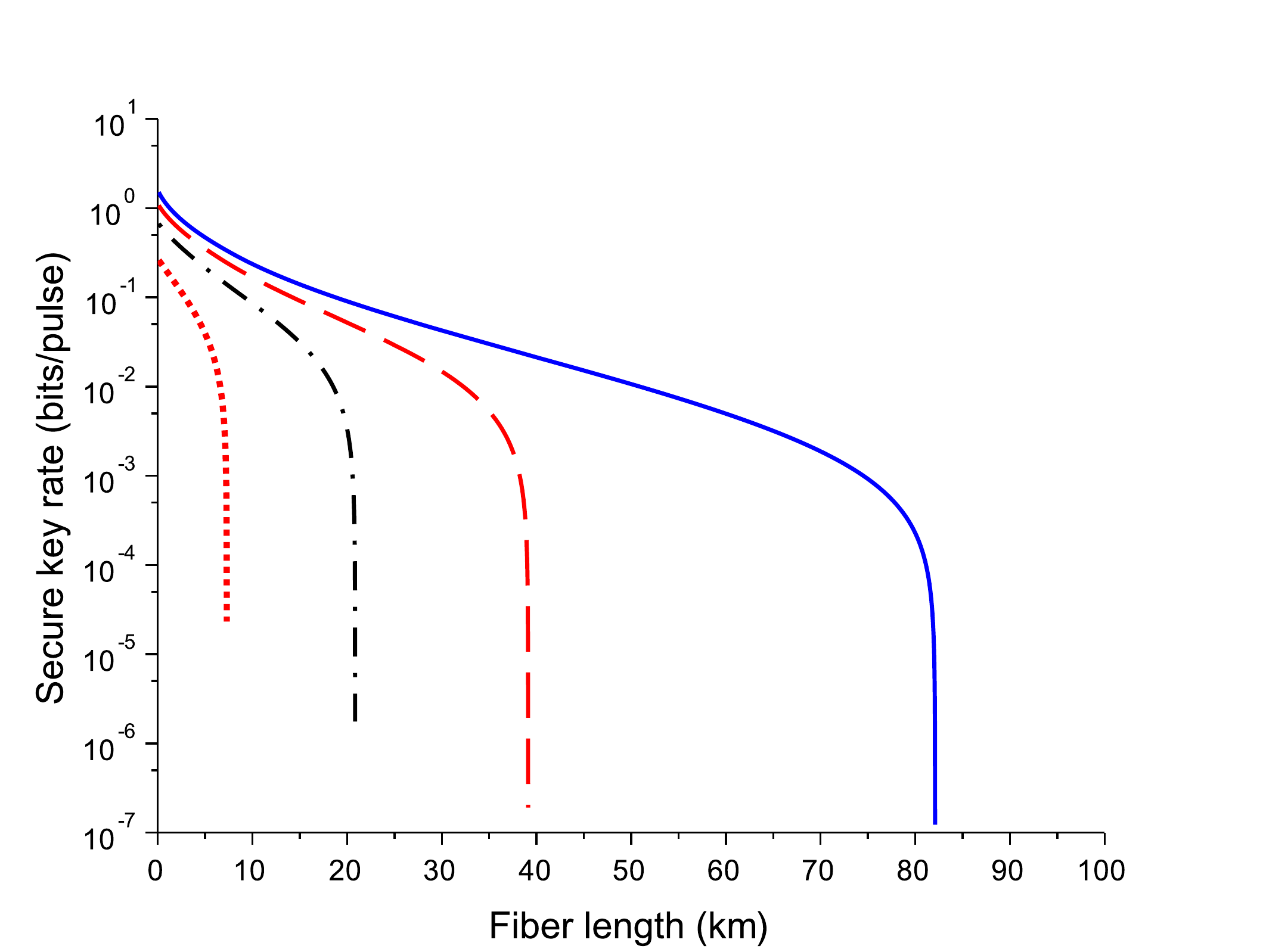}
	\captionsetup{justification=raggedright,
					singlelinecheck=false }
	\caption{Simulation results of the secure key rate for \textit{n}=2 (solid), \textit{n}=5 (dash), \textit{n}=10 (dash dot) and \emph{n}=20 (dot). Simulation parameters: $\gamma=0.2$ dB/km; $\varepsilon_0=0.01$; $\upsilon_{el}=0.1$; $\eta_D=0.5$; $f=0.95$.}
	\label{fig:3}
\end{figure}

\begin{figure}[t]
	\includegraphics[width=.5\textwidth]{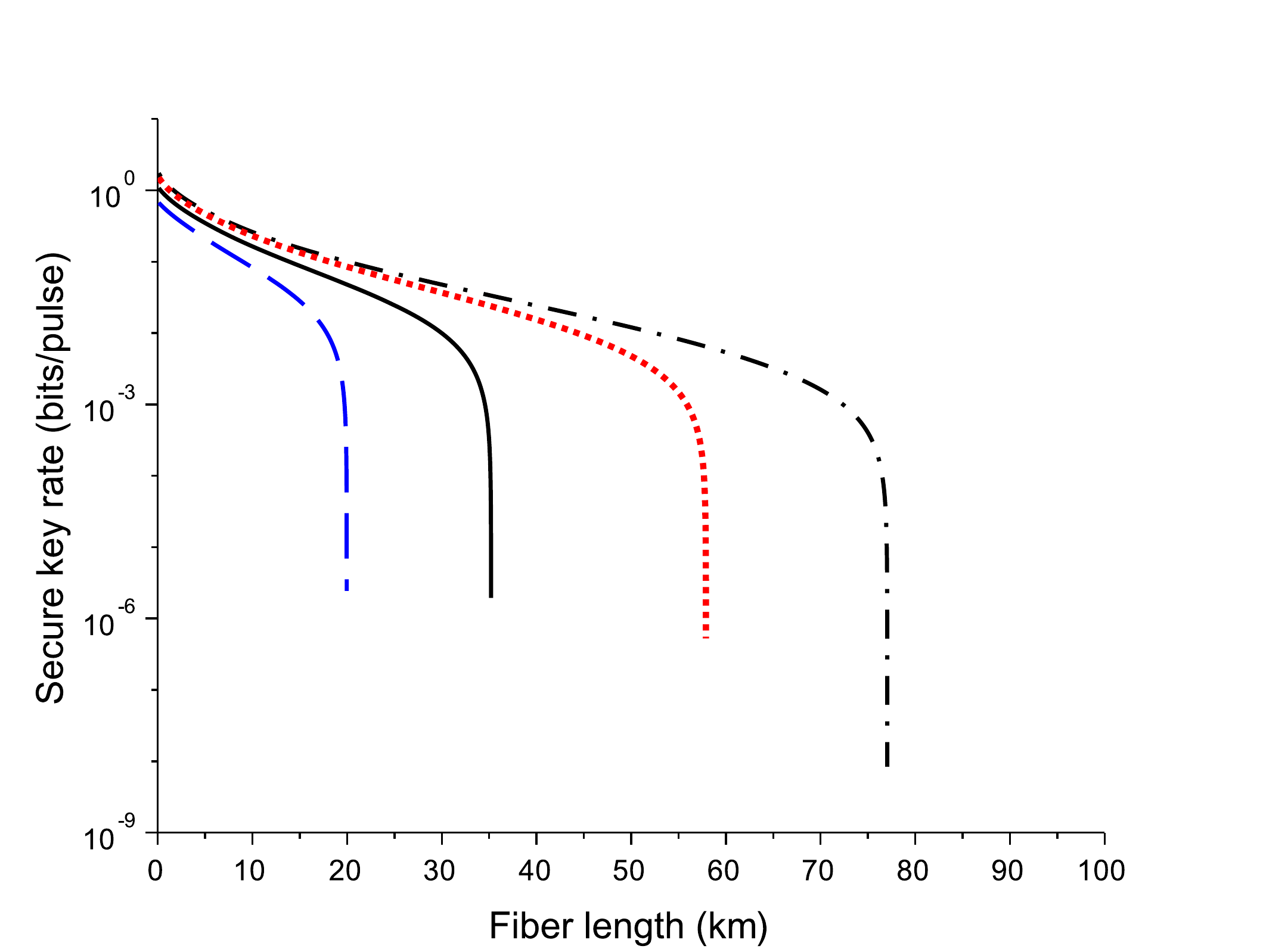}
	\captionsetup{justification=raggedright,
					singlelinecheck=false }
	\caption{Simulation results of the secure key rate for $n=10$ (dash dot), $n=20$ (dot), $n=50$ (solid) and $n=100$ (dash). Simulation parameters: $\gamma=0.2$ dB/km; $\varepsilon_0=0.001$; $\upsilon_{el}=0.1$; $\eta_D=0.5$; $f=0.95$.}
	\label{fig:4}
\end{figure}

\bes\label{eq11} \lambda_{3,4}^2=\frac{1}{2} \left[ C\pm \sqrt{C^2-4D} \right], \ees
where
\begin{equation}
\begin{split}
C=\dfrac{1}{(T_1(V+\chi_{tot}))^2} [ A\chi_{het}^2+B+1+2\chi_{het} \\
( V\sqrt{B}+T_1(V+\chi_{line})) +2T_1(V^2-1)],
\end{split}
\end{equation}

\bes\label{eq12}D=\left( \dfrac{V+\sqrt{B}\chi_{het}}{T_1(V+\chi_{tot})} \right) ^2. \ees 
\bes\label{eq13} \lambda_5=1. \ees

Simulation parameters are summarized as follows:
$\gamma=0.2$ dB/km, $\varepsilon_0=0.01$, $\upsilon_{el}=0.1$, $\eta_D=0.5$, and $f=0.95$. The modulation variance $V_A$ is numerically optimized at different fiber lengths. Note that in (1), when $V_A$ increases, both the mutual information $I_{AB}$ and Eve's information $\chi_{BE}$ will increase. In the ideal case (no excess noise and the efficiency of the reconciliation algorithm $f=1$), a larger modulation variance $V_A$ always leads to a higher secure key rate, so the optimal value of $V_A$ would be infinite. When we take into account system imperfections and non-unity reconciliation efficiency, $\chi_{BE}$ can increase faster than $f I_{AB}$ when $V_A$ is above a certain value. This leads to a finite optimal modulation variance. It is a common practice to numerically search for the optimal value of $V_A$. In this paper, for simplicity, we assume all the players use the same $V_A$. The secure key rate could be further improved by optimizing the modulation variance for each player separately.

In Fig.3 we present the relations of the secure key rate and the fiber length at different numbers of players \textit{n}=2, 5, 10, and 20. As shown in Fig.3, the QSS protocol can be conducted over tens of kilometers of telecom fiber with a moderate number of players. The performance can be further improved by reducing the excess noise contributed by each player. Fig.4 shows the simulation results when $\varepsilon_0=0.001$ and \textit{n}=10, 20, 50, and 100: the QSS protocol can be conducted over $20$ km with 100 players.

\section{Discussion}
\label{sec:4}

Comparing with the previous single qubit sequential QSS scheme \cite{Schmid05}, the CV-QSS proposed here is naturally resilient to Trojan horse attacks: the encoding modulators within the secure stations cannot be reached by the probing signals from external players or the eavesdropper. Furthermore, by using highly asymmetric beam splitters, the additional loss introduced by each player can be extremely small. This opens the door to large-scale implementations. As in the case of single qubit sequential QSS which can be easily changed into a configurable multiuser QKD network \cite{Grice15}, it should be straightforward to implement CV-QKD based on the proposed CV-QSS design. Below we will address a few practical issues.

In Section 2, we have implicitly assumed that all the participants share a phase reference. This allows them to prepare quantum states and perform homodyne detection in the same reference frame. One immediate question is how to establish such a phase reference in practice. One possible solution is the pilot-aided phase recovery scheme proposed in CV-QKD \cite{Qi15, Soh15, Qi17}. The basic idea is that the first player generates a \emph{classical} phase reference pulse using the same laser for quantum state generation. After applying a suitable multiplexing scheme (time, frequency, polarization, or a combination of them), the phase reference pulse propagates through the same optical path as the quantum signal. Each player down the link (and also the dealer) splits out a suitable portion of the phase reference pulse and interferes it with the local laser. This allows each player (and the dealer) to determine the phase difference $\phi_k$ between the local phase frame and that of the first player. After the quantum transmission stage, the $n-1$ players and the dealer first correct the raw data by performing rotation $x_k'=x_k \cos\phi_k-p_k \sin\phi_k$; $p_k'=x_k \sin\phi_k+p_k \cos\phi_k$, then they proceed with the remaining steps of the protocol. This phase recovery scheme has been successfully demonstrated in CV-QKD \cite{Qi15, Soh15, Wang17, Kleis17}.

In practice, the above phase recovery scheme cannot be implemented perfectly due to system imperfections. The additional excess noise contributed by each player can be described by $\varepsilon_p=V_A \delta$, where $\delta$ is the phase noise variance (in units of $\rm{rad^2}$) at each player's station \cite{Qi15}. This additional noise should be added into $\varepsilon_0$ in (3). In \cite{Qi182}, a phase noise of $10^{-3}$ was demonstrated experimentally using the scheme proposed in \cite{Marie17}. We expect that a phase noise of $10^{-4}$ could be achieved by further improving the system. Fig.5 shows the simulation results for the case of $n=20$ players at three different phase noise levels: $\delta=0, 10^{-4}, 10^{-3}$. Even in the case of $\delta=10^{-3}$, a reasonable performance can still be achieved. 

\begin{figure}[t]
	\includegraphics[width=.5\textwidth]{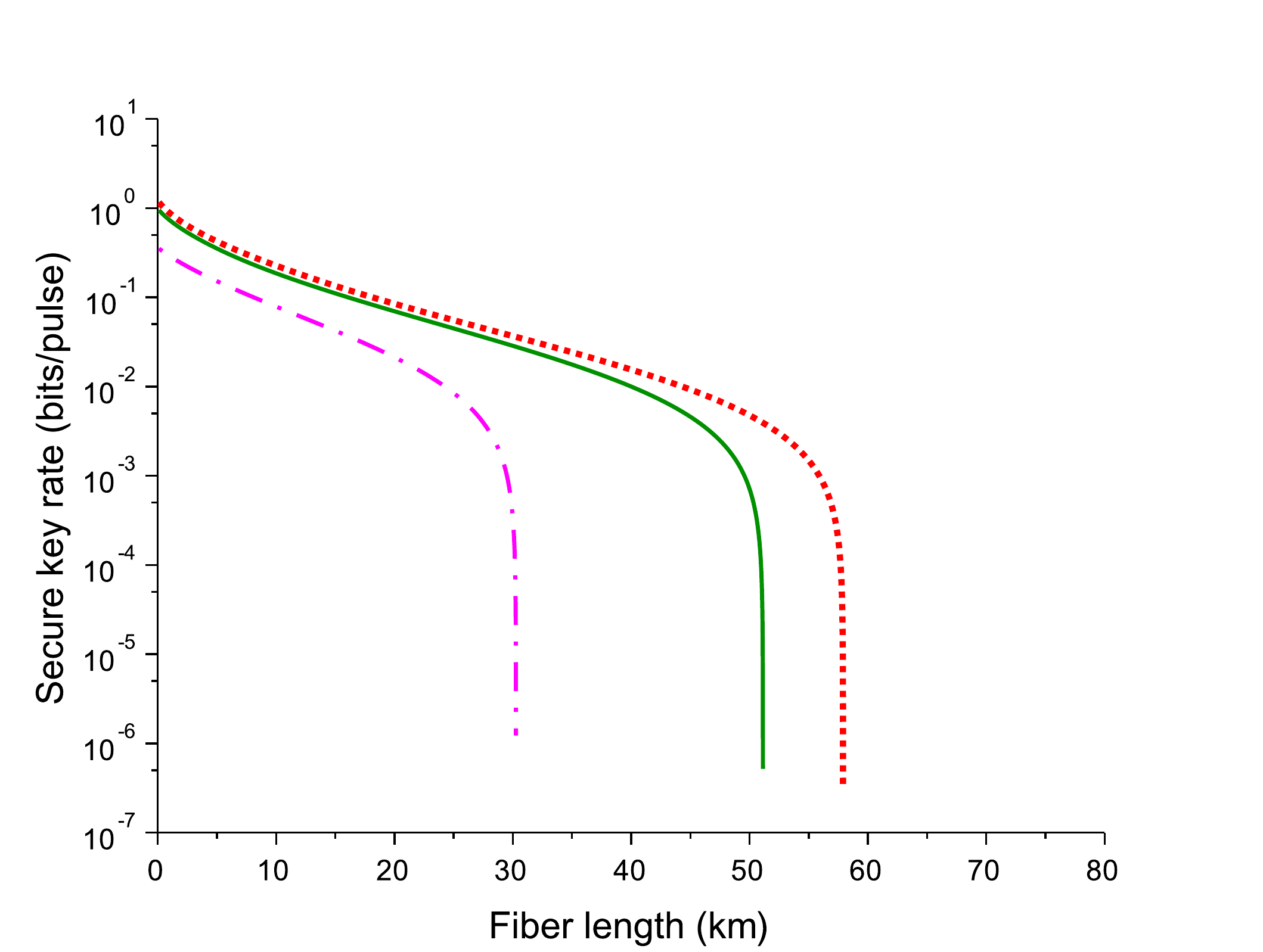}
	\captionsetup{justification=raggedright,
					singlelinecheck=false }
	\caption{Secure key rates in the presence of phase noise for $n=20$.  Simulation results of the secure key rate for phase noise variance $\delta=0$ (dot), $\delta=10^{-4}$ (solid) and $\delta=10^{-3}$ (dash dot). Simulation parameters: $\gamma=0.2$ dB/km; $\varepsilon_0=0.001+V_A \delta$; $\upsilon_{el}=0.1$; $\eta_D=0.5$; $f=0.95$.}
	\label{fig:5}
\end{figure}

The CV QSS protocol proposed in this paper is based on the GMCS QKD, which requires each player to generate Gaussian distributed random numbers and to actively modulate the output of a local laser using phase and amplitude modulators. An alternative passive scheme based on a thermal source has been proposed to simplify the state preparation process in CV-QKD \cite{Qi18}. Such a scheme can also be applied in the proposed CV-QSS protocol. In this case, at each player's station, the phase and amplitude measurements can be carried out with high precision on the portion of the state that is transmitted through the asymmetric beam splitter, rather than on the weaker portion coupled into the quantum channel.

As we noted in Section 2,  in this paper we assume that the dealer performs the homodyne detection. This arrangement allows us to apply the standard security proof of CV-QKD and employ the trusted detector noise model. Can we allow any participant in Fig.2 to be the dealer? One trivial solution is to let each participant have both source and detector. The one chosen as the dealer performs measurement while the others prepare quantum states. This solution requires modifications in the quantum transmission stage and needs complicated system designs and network re-routing. Can we achieve the same goal by only changing the postprocessing procedures? Imagining that after the quantum stage, $P_2$ in Fig.2 decides to be the dealer. $P_2$ could carry out the remaining steps of the protocol as outlined in Section 2, with help from the other participants. More specifically, $P_2$ needs to estimate the potential QKD key rate with each player under the assumption that all the other players are dishonest. There are cases when the two \emph{trusted} QKD parties prepared quantum states while the measurement was conducted by a dishonest player, a scenario as in measurement-device-independent (MDI) QKD \cite{Lo12}. In these cases, the security proof and key rate formulas developed in CV MDI-QKD \cite{Pirandola15, Li14, Ma14} could be applied directly. We remark that the existing schemes of CV MDI-QKD require a highly efficient homodyne detector and are more sensitive to channel losses. We leave the feasibility of CV-QSS based on CV MDI-QKD for future study. 

In summary, we propose a CV-QSS protocol based on practical laser sources and homodyne detectors, which is intrinsically resilient to Trojan horse attacks. By connecting the CV-QSS to CV-QKD, we prove its security against both eavesdroppers and dishonest players in the presence of high channel loss. 

This work was performed at Oak Ridge National Laboratory (ORNL), operated by UT-Battelle for the U.S. Department of Energy (DOE) under Contract No. DE-AC05-00OR22725. The authors acknowledge supports from DOE's Cybersecurity for Energy Delivery Systems Program, Technology Commercialization Fund and ORNL Technology Transfer and Economic Development (Partnerships) Royalty Funds.


\begin{thebibliography}{00}

\bibitem{Shamir79}
A. Shamir, How to share a secret, Commun. ACM \textbf{22}, 612 (1979).

\bibitem{Blakley79}
G. R. Blakley, Safeguarding cryptographic keys,\emph{in Proceedings of the National Computer Conference} (AFIPSPress,1979),Vol. 48, pp. 313–317 

\bibitem{Gisin02}
N. Gisin, G. Ribordy, W. Tittel, and H. Zbinden, Quantum cryptography, Rev. Mod. Phys. \textbf{74}, 145 (2002).

\bibitem{Scarani09}
V. Scarani, H. Bechmann-Pasquinucci, N. J. Cerf, M. Du\v{s}ek, N. L\"{u}tkenhaus, and M. Peev, The security of practical quantum key distribution, Rev. Mod. Phys. \textbf{81}, 1301 (2009).

\bibitem{Lo14}
H.-K. Lo, M. Curty, and K. Tamaki, Secure quantum key distribution, Nat. Photon. \textbf{8}, 595 (2014).

\bibitem{Diamanti16}
E. Diamanti, H.-K. Lo, B. Qi, and Z. Yuan, Practical challenges in quantum key distribution, npj Quantum Inf. \textbf{2}, 16025 (2016).

\bibitem{Hillery99}
M. Hillery, V. Bu\v{z}ek, and A. Berthiaume, Quantum secret sharing, Phys. Rev. A \textbf{59}, 1829 (1999).

\bibitem{Karlsson99}
A. Karlsson, M. Koashi, and N. Imoto, Quantum entanglement for secret sharing and secret splitting, Phys. Rev. A \textbf{59}, 162 (1999).

\bibitem{Tittel01}
W. Tittel, H. Zbinden, and N. Gisin, Experimental demonstration of quantum secret sharing, Phys. Rev. A \textbf{63}, 042301 (2001).

\bibitem{Xiao04}
L. Xiao, G. Lu Long, F.-G. Deng, and J.-W. Pan, Efficient multiparty quantum-secret-sharing schemes, Phys. Rev. A \textbf{69}, 052307 (2004).

\bibitem{Chen05}
Y.-A. Chen, A.-N. Zhang, Z. Zhao, X.-Q. Zhou, C.-Y. Lu, C.-Z. Peng, T. Yang, and J.-W. Pan, Experimental Quantum Secret Sharing and Third-Man Quantum Cryptography, Phys. Rev. Lett. \textbf{95}, 200502 (2005).

\bibitem{Lance04}
A. M. Lance, T. Symul, W. P. Bowen, B. C. Sanders, and P. K. Lam, Tripartite Quantum State Sharing, Phys. Rev. Lett. \textbf{92}, 177903 (2004).

\bibitem{Zhou18}
Y. Zhou, J. Yu, Z. Yan, X. Jia, J. Zhang, C. Xie, and K. Peng, Quantum Secret Sharing Among Four Players Using Multipartite Bound Entanglement of an Optical Field, Phys. Rev. Lett. \textbf{121}, 150502 (2018).

\bibitem{Lau2013}

H.-K. Lau and C. Weedbrook, Quantum secret sharing with continuous-variable cluster states, Phys. Rev. A \textbf{88}, 042313 (2013).

\bibitem{Kogias17}
I. Kogias, Y. Xiang, Q. Y. He, and G. Adesso, Unconditional security of entanglement-based continuous-variable quantum secret sharing, Phys. Rev. A \textbf{95}, 012315 (2017).

\bibitem{Schmid05}
C. Schmid, P. Trojek, M. Bourennane, C. Kurtsiefer, M. Żukowski, and H. Weinfurter, Experimental Single Qubit Quantum Secret Sharing, Phys. Rev. Lett. \textbf{95}, 230505 (2005).

\bibitem{BB84}
C. H. Bennett and G. Brassard, \emph{in Proceedings of IEEE International Conference on Computers, Systems
and Signal Processing} (IEEE Press, New York, 1984), pp. 175-179.

\bibitem{Phoenix95}
S. Phoenix, S. Barnett, P.D. Townsend, and K. Blow, Multi-user quantum cryptography on optical networks, J. Mod. Opt. \textbf{42}, 1155–1163 (1995).

\bibitem{Grice15}
W. P. Grice, P. G. Evans, B. Lawrie, M. Legre, P. Lougovski, W. Ray, B. P. Williams, B. Qi, and A. M. Smith, Two-Party secret key distribution via a modified quantum secret sharing protocol, Opt. Express \textbf{23}, 7300 (2015).

\bibitem{He07}
G. P. He, Comment on Experimental single qubit quantum secret sharing, Phys. Rev. Lett. \textbf{98}, 028901 (2007).

\bibitem{Schmid07}
C. Schmid, P. Trojek, M. Bourennane, C. Kurtsiefer, M. Zukowski, H. Weinfurter, Reply to Comment on `Experimental Single Qubit Quantum Secret Sharing', Phys. Rev. Lett. \textbf{98}, 028902 (2007).

\bibitem{He10}
G. P. He and Z. D. Wang, Single qubit quantum secret sharing with improved security, Quant. Inf. Comput. 10, \textbf{28} (2010).

\bibitem{Muller97}
A. Muller, T. Herzog, B. Huttner, W. Tittel, H. Zbinden, and N. Gisin, ``Plug and Play'' Systems for Quantum Cryptography, Appl. Phys. Lett. \textbf{70}, 793 (1997).

\bibitem{zhao08}
Y. Zhao, B. Qi, and H.-K. Lo, Quantum Key Distribution with an Unknown and Untrusted Source, Phys. Rev. A \textbf{77}, 052327 (2008).

\bibitem{zhao10}
Y. Zhao, B. Qi, H.-K. Lo, and L. Qian, Security Analysis of an Untrusted Source for Quantum Key Distribution: Passive Approach, New J. Phys. \textbf{12}, 023024 (2010).

\bibitem{GMCSQKD}
F. Grosshans, G. V. Assche, J. Wenger, R. Brouri, N. J. Cerf, and Ph. Grangier, Quantum key distribution using gaussian-modulated coherent states, Nature \textbf{421}, 238 (2003).

\bibitem{note1}
Other methods could be used to calibrate channel transmittance, for example by transmitting strong laser pulses.

\bibitem{Diamanti15}
E. Diamanti and A. Leverrier, Distributing Secret Keys with Quantum Continuous Variables: Principle, Security and Implementations, Entropy \textbf{17}, 6072 (2015).

\bibitem{Tomamichel11}
M. Tomamichel and R. Renner, Uncertainty Relation for Smooth Entropies, Phys. Rev. Lett. \textbf{106}, 110506 (2011).

\bibitem{Walk16}
N. Walk, S. Hosseini, J. Geng, O. Thearle, J. Y. Haw, S. Armstrong, S. M. Assad, J. Janousek, T. C. Ralph, T. Symul, H. M. Wiseman, and P. K. Lam, Experimental demonstration of Gaussian protocols for one-sided device-independent quantum key distribution, Optica \textbf{3}, 634 (2016).

\bibitem{Williams2019}
B. P. Williams, J. M. Lukens, N. A. Peters, B. Qi, and W. P. Grice, Quantum Secret Sharing with Polarization-Entangled Photon Pairs, Phys. Rev. A \textbf{99}, 062311 (2019).

\bibitem{Qi15}
B. Qi, P. Lougovski, R. Pooser, W. Grice, and M. Bobrek, Generating the Local Oscillator ``Locally" in Continuous-Variable Quantum Key Distribution Based on Coherent Detection, Phys. Rev. X \textbf{5}, 041009 (2015).

\bibitem{Lodewyck07}
J. Lodewyck, M. Bloch, R. Garc\'{i}a-Patr\'{o}n, S. Fossier, E. Karpov, E. Diamanti, T. Debuisschert, N. J. Cerf, R. Tualle-Brouri, S. W. McLaughlin, and Ph. Grangier, Quantum key distribution over 25km with an all-fiber continuous-variable system, Phys. Rev. A \textbf{76}, 042305 (2007).

\bibitem{Qi07}
B. Qi, L.-L. Huang, L. Qian, and H.-K. Lo, Experimental study on the Gaussian-modulated coherent-state quantum key distribution over standard telecommunication fibers, Phys. Rev. A \textbf{76}, 052323 (2007).

\bibitem{Jouguet13}
P. Jouguet, S. Kunz-Jacques, A. Leverrier, Ph. Grangier, and E. Diamanti, Experimental demonstration of long-distance continuous-variable quantum key distribution, Nat. Photonics \textbf{7}, 378 (2013).

\bibitem{Huang16}
D. Huang, P. Huang, D. Lin, and G. Zeng, Long-distance continuous-variable quantum key distribution by controlling excess noise, Sci. Rep. \textbf{6}, 19201 (2016).

\bibitem{ZLC17}
Y.-C. Zhang, Z. Li, Z. Chen, et al., Continuous-variable QKD over 50km commercial fiber, Quantum Sci. Technol. \textbf{4}, 035006 (2019).

\bibitem{Fossier09}
S. Fossier, E. Diamanti, T. Debuisschert, R. TualleBrouri, and Ph. Grangier, Improvement of continuous-variable quantum key distribution systems by using optical preamplifiers, J. Phys. B \textbf{42}, 114014 (2009).

\bibitem{Soh15}
D. B. S. Soh, C. Brif, P. J. Coles, N. L\"{u}tkenhaus, R. M. Camacho, J. Urayama, and M. Sarovar, Self-Referenced Continuous-Variable Quantum Key Distribution Protocol, Phys. Rev. X \textbf{5}, 041010 (2015).

\bibitem{Qi17}
B. Qi, Pilot-aided feedforward data recovery in optical coherent communications, US Patent 9768885 (2017).

\bibitem{Wang17}
T. Wang, P. Huang, Y. Zhou, W. Liu, H. Ma, S. Wang, and G. Zeng, High key rate continuous-variable quantum key distribution with a real local oscillator, Opt. Express \textbf{26}, 2794 (2018).

\bibitem{Kleis17}
S. Kleis, M. Rueckmann, and C. G. Schaeffer, Continuous variable quantum key distribution with a real local oscillator using simultaneous pilot signals, Opt. Lett. \textbf{42}, 1588 (2017).

\bibitem{Qi182}
B. Qi and C. C. W. Lim, Noise analysis of simultaneous quantum key distribution and classical communication scheme using a true local oscillator, Phys. Rev. Appl. \textbf{9}, 054008 (2018).

\bibitem{Marie17}
A. Marie and R. All\'{e}aume, Self-coherent phase reference sharing for continuous-variable quantum key distribution, Phys. Rev. A \textbf{95}, 012316 (2017).

\bibitem{Qi18}
B. Qi, P. Evans, and W. Grice, Passive state preparation in the Gaussian-modulated coherent-states quantum key distribution, Phys. Rev. A \textbf{97}, 012317 (2018).

\bibitem{Lo12}
H.-K. Lo, M. Curty, and B. Qi, Measurement-Device-Independent Quantum Key Distribution, Phys. Rev. Lett. \textbf{108}, 130503 (2012).

\bibitem{Pirandola15}
S. Pirandola, C. Ottaviani, G. Spedalieri, C. Weedbrook, S. L. Braunstein, S. Lloyd, T. Gehring, C. S. Jacobsen, and U. L. Andersen, High-Rate Measurement-Device Independent Quantum Cryptography, Nat. Photonics \textbf{9}, 397 (2015).

\bibitem{Li14}
Z. Li, Y.-C. Zhang, F. Xu, X. Peng, and H. Guo, Continuous-Variable Measurement-Device-Independent Quantum Key Distribution, Phys. Rev. A \textbf{89}, 052301 (2014).

\bibitem{Ma14}
X.-C. Ma, S.-H. Sun, M.-S. Jiang, M. Gui, and L.-M. Liang, Gaussian-Modulated Coherent-State Measurement-Device Independent Quantum Key Distribution, Phys. Rev. A \textbf{89}, 042335 (2014).

\end{thebibliography}
\end{document}